\documentstyle[12pt,epsfig]{article}

   \font\tenmsb=msbm10 scaled\magstep 1
   \font\sevenmsb=msbm7 scaled \magstep 1
   \font\faivemsb=msbm5 scaled \magstep 1
\newfam\msbfam
      \textfont\msbfam=\tenmsb
      \scriptfont\msbfam=\sevenmsb
      \scriptscriptfont\msbfam=\faivemsb
\def\Bbb#1{{\fam\msbfam #1}}
\font\tengothic=eufm10 scaled\magstep 1
\font\sevengothic=eufm7 scaled\magstep 1
\newfam\gothicfam
      \textfont\gothicfam=\tengothic
      \scriptfont\gothicfam=\sevengothic

\newcommand{\be}{\begin{equation}}
\newcommand{\ee}{\end{equation}}

\newcommand{\bt}{\beta}
\newcommand{\vp}{\varphi}
\newcommand{\ep}{\varepsilon}
\newcommand{\al}{\alpha}
\newcommand{\ra}{\rightarrow}

\newcommand{\Gm}{\Gamma}

\newcommand{\lbd}{\lambda}

\begin{document}

\begin{center}

{\Large{\bf Method of self-similar factor approximants} \\ [5mm]
V.I. Yukalov$^{1,2}$ and E.P. Yukalova$^3$} \\ [3mm]

{\it $^1$Bogolubov Laboratory of Theoretical Physics, \\
Joint Institute for Nuclear Research, Dubna 141980, Russia, \\ [2mm]

$^2$Department of Technology and Economics, \\
Swiss Federal Institute of Technology, Z\"urich CH-8032, Switzerland, \\ 
[2mm] 

$^3$Laboratory of Information Technologies, \\
Joint Institute for Nuclear Research, Dubna 141980, Russia}

\end{center}

\vskip 3cm

\begin{abstract}

The method of self-similar factor approximants is completed by defining 
the approximants of odd orders, constructed from the power series with the 
largest term of an odd power. It is shown that the method provides good 
approximations for transcendental functions. In some cases, just a few 
terms in a power series make it possible to reconstruct a transcendental 
function {\it exactly}. Numerical convergence of the factor approximants 
is checked for several examples. A special attention is paid to the 
possibility of extrapolating the behavior of functions, with arguments 
tending to infinity, from the related asymptotic series at small arguments.
Applications of the method are thoroughly illustrated by the examples of 
several functions, nonlinear differential equations, and anharmonic 
models.

\end{abstract}

\vskip 2cm


\newpage

\section{Introduction}

The problem of reconstructing functions from their asymptotic series is 
widely known to be of great importance for a variety of applications [1]. 
Probably, the most often used technique allowing for such a reconstruction 
is that based on Pad\'e approximants [2], though there also exist other more 
complicated methods [3].

Recently, a novel approach has been advanced for reconstructing functions 
from the related asymptotic series, called the method of {\it self-similar 
factor approximants} [4--6]. The form of these approximants is derived by 
means of the self-similar approximation theory [7--15]. The factor 
approximants were shown [4--6] to be more general and more accurate than 
Pad\'e approximants. However, several points in the method of factor 
approximants [4--6] have not been investigated.

The most important point, which has remained unclear, is how to construct 
the factor approximants of {\it odd orders}? The problem is that the standard 
procedure [4--6] requires that the generic series, the factor approximants 
are built from, be of even order. So that the large amount of information, 
contained in the odd orders of the series, could not be properly used. Here 
we advance a uniform approach for treating both odd as well as even orders 
of power series and we demonstrate, by a number of examples, good performance 
of this general method.

Another, rather nontrivial, question is why the factor approximants could 
provide high accuracy for {\it transcendental functions}, obtained from their 
asymptotic series. We give an explanation for this effect, which is unique 
for the resummation methods based on asymptotic series. Recall that Pad\'e 
approximants, having the structure of rational functions, are usually not so 
good for approximating transcendental functions. Moreover, we show that some 
transcendental functions can be reconstructed, by means of the factor
approximants, {\it exactly} from a few terms of an asymptotic series.

In the previous papers [4--6] on the method of self-similar factor 
approximants, we mainly considered the low-order approximants of even order, 
because of which their numerical convergence could not be properly analyzed. 
Now, we possess both even and odd orders of these approximants, and we study 
their numerical convergence for various cases, calculating high-order 
approximants, up to 20-th order.

We pay a special attention to the possibility of extrapolating physically 
motivated functions to the whole range of their variables. We demonstrate 
that the method of self-similar factor approximants can serve as a tool 
for defining the values of functions at infinity from their asymptotic 
expressions near zero.

Summarizing, the main results of the present paper are as follows:

\vskip 2mm

(i) The method of self-similar factor approximants is extended by 
constructing the factor approximants of odd orders, in addition to those of 
even orders.

\vskip 2mm

(ii) It is shown that transcendental functions can be well approximated. 
A unique feature of the method is that some transcendental functions can 
be reconstructed exactly.

\vskip 2mm

(iii) The method allows for the extrapolation of solutions to nonlinear 
differential equations from asymptotically small to finite variables.

\vskip 2mm

(iv) The characteristics of quantum anharmonic models can be extrapolated 
from the region of asymptotically small coupling parameter to the whole 
range of the latter, including the case of the coupling parameter tending 
to infinity.

\vskip 2mm

(v) For all considered examples, the factor approximants of high orders 
are calculated, demonstrating the existence of numerical convergence.

\section{Self-similar factor approximants}

Let us be interested in defining a real function $f(x)$ of a real variable 
$x\in\Bbb{R}$. The extension of the method to complex-valued functions of 
complex variables is also possible, but, first, let us consider a slightly 
simpler case of real functions and variables. Suppose, we know only the 
behavior of the function at asymptotically small values of $x\ra0$, where 
we can get the sequence $\{ f_k(x)\}$ of the expansions
\be
\label{1}
f_k(x) = \sum_{n=0}^k a_n x^n \, ,
\ee
with $k=0,1,2,\ldots$. The sequence, generally, can be divergent. It is known 
[1] that a function $f(x)$, analytic in the vicinity of $x=0$, uniquely defines 
its asymptotic series. The converse is not always true. But in what follows,
we consider the situation, when there is a one-to-one correspondence between
the function and its asymptotic series, since solely then the problem of 
reconstructing functions from their expansions acquires sense.

Without the loss of generality, we may assume that in expansion (1),
\be
\label{2}
a_0 = f_k(0) = f_0(x) = 1\; .
\ee
Really, if instead of expansion (1), we would have
$$
f^{(k)}(x) = f^{(0)}(x) \sum_{n=0}^k a_n x^n\; ,
$$
with a given function $f^{(0)}(x)$, then we could immediately return to Eq. (1)
defining
$$
f_k(x) \equiv \frac{f^{(k)}(x)}{f^{(0)}(x) a_0}\; .
$$

The self-similar factor approximants of even orders $k=2p=2,4,6,\ldots$ are 
given [4--6] by the form
\be
\label{3}
f_{2p}^*(x) = \prod_{i=1}^p \left ( 1 + A_i x\right )^{n_i}\; ,
\ee
whose parameters $A_i$ and $n_i$ are obtained from the re-expansion procedure, 
that is, by expanding Eq. (3) in powers of $x$ and comparing the results with 
the given expansion (1), equating the like-order terms. This 
accuracy-through-order procedure yields the set of equations
\be
\label{4}
\sum_{i=1}^p n_i A_i^n = B_n \qquad (n=1,2,\ldots,2p)\; ,
\ee
with the right-hand sides
\be
\label{5}
B_n \equiv \frac{(-1)^{n-1}}{(n-1)!}\; \lim_{x\ra 0}\;
\frac{d^n}{dx^n}\; \ln f_k(x) \; .
\ee
In each approximation, the parameters $A_i$, $n_i$, and $B_n£$, of course,
depend  on the approximation number $k=2p$. However, in order to avoid too 
cumbersome notation, we do not mark explicitly this dependence, which is 
assumed to be evident.

Each factor in form (3) contains two parameters, $A_i$ and $n_i$. This is why 
these approximants could be straightforwardly defined only for the even-order 
expansions $f_{2p}(x)$. But how should we proceed having odd-order expansions 
(1) with $k=2p+1=1,3,5,\ldots$? For the latter, we could write the form 
\be
\label{6}
f^*_{2p+1}(x) = \prod_{i=1}^{p+1} \left ( 1 + A_i x \right )^{n_i}\; ,
\ee
with the parameters satisfying the set of equations
\be
\label{7}
\sum_{i=1}^{p+1} n_i A_i^n = B_n \qquad \left ( n=1,2,\ldots, 2p+1
\right ) \; .
\ee
But the problem is that form (6) contains $2p+2$ unknown $A_i$ and $n_i$, 
while only $2p+1$ equations of set (7) are available. One equation is 
lacking.

To overcome this problem, let us notice that expression (6) is invariant 
under the scaling transformation
\be
\label{8}
x \ra \lbd x \; , \qquad A_i \ra \lbd^{-1} A_i \; .
\ee
Then, taking $\lbd \ra A_1^{-1}$ and using the renotation $A_i/A_1\ra A_i$, 
we come to the same form (6), but with 
\be
\label{9}
A_1 = 1 \qquad (k=2p+1) \; .
\ee
Complementing the set of $2p+1$ equations (7) by the scaling condition (9), 
we get $2p+2$ equations for $2p+2$ unknowns. In this way, we now can construct
the factor approximants of odd orders.

If the sought function $f(x)$ can be reduced to the factor form $f_k^*(x)$, 
then it is obvious that this function can be reconstructed exactly for all 
approximants, starting with the given order $k$, since both $f(x)$ and 
$f_k^*(x)$ yield the same series of the $k$-th order. Thus, there exists 
a class of functions $\{ f_k^*(x)\}$, having the form of $f_k^*(x)$, which 
can be reconstructed exactly from their asymptotic series by means of the 
factor approximants. This class of exactly reproduceable functions can be 
noticeably extended by including those functions that can be defined as 
limits of the form $f_k^*(x)$ with respect to the values of some $A_i$ 
and $n_i$ tending to either zero or infinity. Keeping in mind such limits, 
we shall denote them as $\lim_{\{ A_i,n_i\}}$. All functions, which can 
either be reduced to the factor form $f_k^*(x)$ or obtained from the 
latter by means of limiting procedures with respect to their parameters, 
form the factor 
class
\be
\label{10}
\Bbb{F}\equiv \left \{ f_k^*(x), \lim_{\{ A_i,n_i\} }
f_k^*(x) \right \} \; .
\ee
As is clear, a function can be exactly reproduced by the factor approximants 
if and only if it pertains to the factor class (10). This is a rather wide 
class, including rational as well as irrational functions. In the following 
section, we show that some transcendental functions can also be reproduced 
exactly.

\section{Reconstruction of transcendental functions}

Consider such an entire transcendental function as the exponential
\be
\label{11}
f(x) = e^x = \sum_{n=0}^\infty \; \frac{x^n}{n!} \; .
\ee
From the expansion $f_2(x)=1+x+\frac{1}{2}x^2$, we find the factor 
approximant
\be
\label{12}
f_2^*(x) = \lim_{a\ra 0} (1+ax)^{1/a} = e^x \; .
\ee
Similarly, from the form $f_3(x)$ at $x\ra 0$, we obtain
\be
\label{13}
f_3^*(x) = \lim_{b\ra 0} (1+x)^{b/(1-b)} (1+bx)^{1/b(1-b)} = e^x \; .
\ee
This procedure can be continued to any order, showing that the transcendental 
function $e^x$ is exactly reproduceable by the factor approximants of any 
finite order $k\ge 2$,
\be
\label{14}
f_k^*(x) = e^x \qquad (k\geq 2) \; .
\ee

The hyperbolic functions, such as ${\rm sinh}x$ and ${\rm cosh}x$, can also 
be reconstructed exactly, if one notices that the series, corresponding to 
$2{\rm sinh}x=e^x-e^{-x}$ and $2{\rm cosh}x=e^x+e^{-x}$, can be represented 
as the sums of two series for the exponentials, which, as is shown above, 
are reproduceable exactly. The trigonometric functions $\sin x$ and $\cos 
x$ can also be reproduced exactly, if one treats them as the analytic 
continuations of the related hyperbolic functions to the imaginary axis, 
when $x$ is replaced by $ix$.

The fact that some transcendental functions can be reproduced by factor 
approximants exactly does not imply that all transcendental functions are 
exactly reproduceable. However, the very existence of exactly reproduceable 
transcendental functions explains why other transcendental functions, though 
not exactly reproduceable, nevertheless, can be well approximated by the 
factor approximants.

As an example, let us consider the series for $\sin x$,
\be
\label{15}
\sin x = \sum_{n=0}^\infty \; \frac{(-1)^n}{(2n-1)!} \; x^{2n+1} \; .
\ee
Note that for the alternating series, it is convenient to shift them by a 
constant, so that all factor approximants for the shifted function be sign 
defined, removing this shift afterwards. The actual value of the shifting 
constant plays no role, provided the considered function becomes sign 
defined. In the present case, we, first, consider $\sin x+const$, and 
remove the constant at the end. It is important to emphasize that the factor 
approximants, obtained from the Taylor series, are much more accurate than 
the latter. For an illustration, we show in Fig. 1 the factor approximant 
of the order $k=18$, derived from series (15), and the Taylor series of the 
same order $k=18$, as compared to the exact function $\sin x$. As is seen, 
the factor approximant is essentially more accurate.

Constructing the factor approximants for the tangent,
\be
\label{16}
\tan x = \sum_{n=0}^\infty (-1)^n\; 2^{2n}\; \frac{2^{2n}-1}{(2n)!} \; 
B_{2n} x^{2n-1} \; ,
\ee
where $B_n$ is the Bernoulli number, we get such a good accuracy for the 
approximants of the order $k\geq 4$, that they practically coincide with the 
exact function. Therefore, we concentrate on defining the point of singularity, 
which for $\tan x$ occurs at $x=\pm\pi/2$. For each factor approximant, we 
find the point of its singularity $x_k$ and compare it with the exact value 
$\pi/2$. We consider the positive value $x>0$, since for $x<0$, the 
behavior is similar, being antisymmetric. The accuracy of the approximants 
can be characterized by the percentage error
\be
\label{17}
\ep_k \equiv \frac{x_k-\pi/2}{\pi/2} \cdot 100\% \; .
\ee
The error quickly diminishes with the approximation order, demonstrating 
fast numerical convergence, for instance, $\ep_2=-6.8\%$, $\ep_3=-5.4\%$, 
$\ep_4 = 0.13\%$, $\ep_5=0.096\%$, and $\ep_6 = -0.00035\%$.

Numerical convergence is well seen for different transcendental functions. 
We have considered a variety of the latter. For illustration, we show in 
Fig. 2 how the factor approximants approach the function
\be
\label{18}
f(x) = x +\cos x \; .
\ee

Often transcendental functions appear as solutions of differential 
equations. Let the function $y(t)$ of time $t\geq 0$ be defined by a 
differential equation. The simplest way to construct its solutions in 
the form of the factor approximants is as follows. We may look for the 
solution at small $t\ra 0$, representing it as a series
$$
y(t) \simeq \sum_n a_n t^n \qquad (t\ra 0) \; .
$$
And then, it is straightforward to derive the factor approximants from the 
given series, as is explained in Sec. 2.

Let us consider the nonlinear ordinary differential equation
\be
\label{19}
\frac{d^2 y}{dt^2} + y +y^2 = 0 \; ,
\ee
describing a nonlinear oscillator with the cubic potential
$$
U = \frac{1}{2}\; y^2 + \frac{1}{3}\; y^3  \; .
$$
The solution to this equation (19) corresponds to the oscillations around 
the center ($y=0$, $\dot{y}=0$), where the overdot, as usual, means time 
derivative, during the time interval $0\leq t< t_0$. At the point $t_0$,
depending on initial conditions, the solution diverges as 
$y\propto(t-t_0)^{-2}$. Constructing the solutions to this equation from 
the related time series at $t\ra 0$, as is explained above, we obtain the 
factor approximants $y_k^*(t)$ providing the approximate solutions for longer 
times. In Fig. 3, the factor approximants $y_{14}^*(t)$ and $y_{16}^*(t)$ 
are shown in comparison with the exact numerical solution of Eq. (19) with 
the initial conditions $y(0)=0$ and $\dot{y}(0)=1$. For these initial 
conditions, the divergence occurs at $t_0\cong 6.5$. We have analyzed the 
behavior of high-order approximants $y_k^*(t)$, with $k\geq 15$, finding 
that they provide a good extrapolation from the region of asymptotically 
small times $t\ra 0$ to the temporal interval $0\leq t\leq 5$. 

As another example of nonlinear differential equations, let us take the 
Rayleigh equation
\be
\label{20}
\frac{d^2 y}{dt^2} + y = \ep\; \frac{dy}{dt} -  \frac{\ep}{3}\left (
\frac{dy}{dt}\right )^3 \; ,
\ee
describing an oscillator perturbed by the right-hand side of Eq. (20), 
with $0<\ep\ll 1$. Written in the normal form, Eq. (20) represents a 
two-dimensional dynamical system having one unstable fixed point 
($y=0$, $\dot{y}=0$), with the Jacobian eigenvalues
$J^\pm=(\ep\pm\sqrt{\ep^2 - 4})/2$. By Poincar\'e-Bendixon theorem [16], 
there exists a limit cycle around this unstable fixed point. We look for 
the solution of Eq. (20), with $\ep=0.1$, under the initial conditions 
$y(0)=0$, $\dot{y}(0)=1$. As is described above, we, first, find the 
power-series solution at $t\ra 0$, from which we construct the factor 
approximants $y_k^*(t)$. In Fig. 4, the approximants $y_{18}^*(t)$ and 
$y_{19}^*(t)$ are compared to the exact numerical solution of Eq. (20). 
Again we observe that the factor approximants of the order $k\geq 19$ 
well extrapolate the time-series solution at $t\ra 0$ to the finite 
region $0\leq t\leq 6$.

It is worth noting that the problem of extrapolating time series is of 
high importance for the possibility of predicting the evolution of different 
complex systems such as markets and societies [17]. The examples, considered 
above, demonstrate that the self-similar factor approximants provide us with 
the tool of extrapolating the time series for small times $t\ra 0$ to the 
region of finite times, that is, making it possible to give forecasts, at 
least, for some near future.

\section{Extrapolation for anharmonic models}

Many interacting physical systems have the structure analogous to 
anharmonic models. This is why the latter serve as a touchstone for 
testing different approximation theories. The question to be investigated 
in the present section is as follows. Suppose, we are able to find a 
function of interest only by means of perturbation theory with respect 
to the coupling parameter $g$. We keep in mind here the simple perturbation 
theory in powers of $g$, without invoking any optimization procedures (see 
review-type articles [12,15]). Then, as is well known, the sought function 
has the form of an asymptotic series
\be
\label{21}
f(g) \simeq \sum_n a_n g^n \qquad (g\ra 0) \; ,
\ee
which diverges for any finite $g$. Could we, being based solely on this 
asymptotic expansion, extrapolate by factor approximants the function to the 
whole region of the coupling parameter $g\in[0,\infty)$? It is especially 
interesting whether we could predict the behavior of the function at very 
large couplings $g\ra\infty$. The latter has, clearly, no sense for the 
power series derived for $g\ra 0$. Pad\'e approximants are also not of the 
help for this purpose, since the limit $g\ra\infty$ for a Pad\'e approximant 
$P_{[M/N]}(g)$ is not defined, as
$$
P_{[M/N]}(g) \; \propto \; g^{M-N} \qquad (g\ra \infty)
$$
can tend to infinity, zero, or a constant, depending on $M$ and $N$, so that, 
as $g\ra\infty$,
\begin{eqnarray}
\nonumber
P_{[M/N]}(g) \ra \left \{ 
\begin{array}{ll}
\infty , & M > N \\
const, & M=N \\
0, & M< N \; . \end{array}  \right.
\end{eqnarray}

First, let us consider the so-called partition function of the 
zero-dimensional $\vp^4$ model,
\be
\label{22}
Z(g) = \frac{1}{\sqrt{\pi}} \; \int_{-\infty}^\infty \; e^{-\vp^2-g\vp^4}
d\vp \; .
\ee
The power series (21), corresponding to Eq. (22), possess the coefficients
$$
a_n = \frac{(-1)^n}{\sqrt{\pi} n!}\; \Gm\left ( 2n +
\frac{1}{2}\right ) \; ,
$$
which factorially increase at large $n$, because of which series (21) 
diverges for any finite $g$.

Being based on series (21), we construct, according to the general scheme 
of Sec. 2, the factor approximants $Z^*_k(g)$ and check their accuracy by 
calculating percentage errors
\be
\label{23}
\ep_k(g) \equiv \frac{Z^*_k(g)-Z(g)}{Z(g)}\; \cdot 100\%
\ee
with respect to the exact value (22). The results for different approximation 
orders and coupling parameters are presented in Table 1. This shows that the 
even-order factor approximants demonstrate the uniform numerical convergence 
for all $g$, and the errors of the odd-order approximants oscillate in the 
frame of the errors of the even-order approximants.

At large $g$, the partition function (22) has the asymptotic behavior
\be
\label{24}
Z(g) \simeq 1.022765\; g^{-1/4} \qquad (g\ra\infty) \; .
\ee
It is interesting whether the factor approximants $Z_k^*(g)$, based solely 
on the power expansion (21) at small $g\ra 0$, could, nevertheless, catch the 
behavior of $Z(g)$ at large $g\ra\infty$. The most nontrivial would be to 
predict the noninteger index $-1/4$ in the asymptotic form (24). For the found
factor approximants, we study the limiting expression
\be
\label{25}
Z_k^*(g) \simeq c_k g^{-\al_k} \qquad (g\ra\infty) \; 
\ee
and define the percentage error of the predicted index $\al_k$ as
\be
\label{26}
\ep_k(g) \equiv \frac{\al_k-0.25}{0.25}\; \cdot 100\% \; .
\ee
In Table 2, we present the parameters of the asymptotic form (25), 
characterizing the behavior of the factor approximants at large coupling 
$g\ra\infty$, together with the percentage error (26) for the index $\al_k$.
Again, we see that the even-order approximants display numerical convergence, 
and the odd-order approximants oscillate around the values of the even-order 
ones. The index $\al_k$, as well as the amplitudes $c_k$, predict the 
large-coupling limit (24) with an error of about $20\%$. This is not as bad, 
if we remember that all calculations are based solely on the weak-couping 
expansion (21), valid for $g\ra 0$, and no additional optimization procedures 
have been involved.

Another standard touchstone for testing new approximations is the calculation
of the ground-state energy for the one-dimensional anharmonic oscillator with 
the Hamiltonian
\be
\label{27}
H = -\; \frac{1}{2}\; \frac{d^2}{dx^2} + \frac{1}{2}\; x^2 + gx^4 \; ,
\ee
in which $x\in(-\infty,\infty)$ and $g\in[0,\infty)$. The weak-coupling 
expansion (21) for the ground-state energy $E(g)$ is known [18,19]. In the 
latter references, one can find the corresponding values of the expansion 
coefficients $a_n$.

Following the standard procedure of Sec. 2, we construct the factor 
approximants $E_k^*(g)$ for the ground-state energy $E(g)$ of the anharmonic 
oscillator with the Hamiltonian (27) and compare the obtained $E_k^*(g)$ with 
the exact numerical data [20]. The percentage errors
\be
\label{28}
\ep_k(g) \equiv \frac{E^*_k(g)-E(g)}{E(g)}\; \cdot 100\%
\ee
of the factor approximants are given in Table 3. Our results show that the 
even-order and odd-order approximants form two sequences $\{ E_{2p}^*(g)\}$ 
and $\{ E_{2p+1}^*(g)\}$ each of which uniformly converges for all $g$.

We again pay a special attention to the ability of factor approximants to 
predict the behavior of the sought function at infinitely large coupling 
parameter $g\ra\infty$. Then the ground-state energy behaves as
\be
\label{29}
E(g) \simeq 0.667986\; g^{1/3} \; .
\ee
We analyze the factor approximants $E_k^*(g)$, obtained on the basis of the 
weak-coupling expansion (21) at $g\ra 0$, in the strong-coupling limit 
$g\ra\infty$ and compare their asymptotic form
\be
\label{30}
E_k^*(g) \simeq b_k g^{\bt_k} \qquad (g\ra\infty)
\ee
with the exact asymptotic behavior (29). The results are presented in 
Table 4, where also the percentage error
$$
\ep_k(\bt) \equiv \frac{\bt_k-1/3}{1/3}\; \cdot 100\% 
$$
of the predicted index $\bt_k$ is shown. The even-order approximants 
demonstrate numerical convergence, and the errors of the odd-order 
approximants fluctuate around those of the even-order approximants. The 
errors of the amplitudes $b_k$ are close to $10\%$ for all $k\geq 2$.

\section{Conclusion}

The method of self-similar factor approximants has been completed by 
defining a general way for constructing both even-order as well as 
odd-order approximants. It is shown that transcendental functions can 
be well approximated by factor approximants. Moreover, some transcendental 
functions can be reconstructed {\it exactly} from just a few terms of their 
asymptotic expansions at small arguments. For all cases considered, the 
even-order approximants demonstrate uniform numerical convergence. The 
accuracy of the odd-order approximants is close to that of the even-order 
ones.

It is shown that the solutions to nonlinear differential equations can 
be extrapolated from the values of asymptotically small variables to the 
finite region of the latter.

A special attention has been paid to the possibility of predicting the 
asymptotic behavior of functions at their arguments tending to infinity 
from the knowledge of their asymptotic expansions at zero. Such an extreme 
extrapolation is shown to be feasible by means of the factor approximants.

Also, we have studied the problem of a possible evaluation of the accuracy of 
the factor approximants, when the exact solution would not be available. For 
this purpose, we have compared, for the cases analyzed above, the quantities 
$(f_k^*-f_{k-1}^*)/2$ and $f_k^*-f$, where $f$ is the known exact numerical 
solution. It turned out that these quantities are close to each other. Hence 
the difference $(f_k^*-f_{k-1}^*)/2$ can serve as a reasonable estimate for 
the accuracy of the $k$-th order factor approximant $f_k^*$.

\newpage

\begin{center}

{\Large{\bf Figure Captions}}

\end{center}

\vskip 5mm

{\bf Fig. 1}. Factor approximant $f_{18}^*(x)$ (dashed line) and the 
Taylor series of the same order $k=18$ (dashed-dotted line), as compared 
to the exact $\sin x$ (solid line).

\vskip 5mm

{\bf Fig. 2}. Factor approximants $f_{8}^*(x)$ (dashed-dotted line), 
$f_{9}^*(x)$ (dotted line), and $f_{10}^*(x)$ (dashed line), as compared 
to the exact function (18) (solid line).

\vskip 5mm

{\bf Fig. 3}. Solutions to the nonlinear oscillator equation (19), with 
the initial conditions $y(0)=0$ and $\dot{y}(0)=1$, obtained from the time 
series at $t\ra 0$ in the form of the factor approximants $y_k^*(t)$. Here 
shown are $y_{14}^*(t)$ (dotted line), $y_{16}^*(t)$ (dashed line), and the 
exact numerical solution (solid line).

\vskip 5mm

{\bf Fig. 4}. Solutions to the Rayleigh equation (20) in the form of the 
factor approximants $y_{18}^*(t)$ (dotted line) and $y_{19}^*(t)$ (dashed 
line), compared to the exact numerical solution (solid line).

\newpage

\begin{center}

{\Large{\bf Table Captions}}

\end{center}

\vskip 5mm

{\bf Table 1}. Percentage errors of the factor approximants $Z_k^*(g)$ for 
the partition function of the anharmonic model (22).

\vskip 5mm

{\bf Table 2}. Amplitudes $c_k$ and indices $\al_k$ for the strong-coupling 
limit (25) of the factor approximants for the partition function (22), together 
with the percentage errors (26) for the predicted indices.

\vskip 5mm

{\bf Table 3}. Percentage errors of the factor approximants $E_k^*(g)$ for
the ground-state energy of the anharmonic oscillator.

\vskip 5mm

{\bf Table 4}. Amplitudes $b_k$ and indices $\bt_k$ for the strong-coupling 
limit (30) of the factor approximants for the ground-state energy of the 
anharmonic oscillator, with the percentage errors $\ep_k(\bt)$ of the
predicted indices $\bt_k$.

\newpage

\begin{center}

{\large{\bf Table 1}}

\vskip 1cm

\begin{tabular}{|c|c|c|c|c|c|} \hline
$k\setminus g$&   0.1    &   1   &  5   & 10   & 100   \\ \hline
 2 &  0.2     &  5.4  &  19  &  27  &  70   \\ 
 3 & -0.07    & -4.6  & -17  & -23  & -41   \\ 
 4 &  0.02    &  2.0  &  9.4 &  15  &  42 \\
 5 & -0.008   & -1.3  & -5.7 & -8.0 & -13 \\ 
 6 &  0.003   &  0.9  &  5.8 &  9.7 &  30 \\
 7 & -0.001   & -0.5  & -2.0 & -2.5 & -1.6 \\
 8 &  0.0006  &  0.5  &  3.9 &  6.9 &  23 \\
 9 & -0.0002  & -0.2  & -0.5 & -0.2 &  3.8 \\ 
10 &  0.0001  &  0.3  &  2.8 &  5.2 &  19 \\
11 & -0.00004 & -0.05 &  0.2 &  0.8 &  6.4 \\
12 &  0.00004 &  0.2  &  2.1 &  4.1 &  16  \\ \hline
\end{tabular}

\newpage

{\large{\bf Table 2}}

\vskip 1cm

\begin{tabular}{|c|c|c|c|} \hline
$k$ & $c_k$  & $\al_k$ &  $\ep_k(\al)$ \\ \hline
 2  &  0.823 &   0.090 &   -64  \\ 
 3  &  0.917 &   0.346 &    38  \\ 
 4  &  0.806 &   0.129 &   -48  \\
 5  &  0.878 &   0.255 &    2.0 \\ 
 6  &  0.806 &   0.148 &   -41  \\
 7  &  0.860 &   0.223 &   -11  \\
 8  &  0.810 &   0.161 &   -36  \\
 9  &  0.850 &   0.209 &   -16  \\ 
 10 &  0.814 &   0.170 &   -32  \\
 11 &  0.845 &   0.202 &   -19  \\
 12 &  0.819 &   0.178 &   -29  \\ 
 13 &  0.842 &   0.199 &   -20  \\
 14 &  0.824 &   0.182 &   -27  \\
 15 &  0.841 &   0.197 &   -21  \\
 16 &  0.828 &   0.187 &   -25  \\
 17 &  0.840 &   0.196 &   -22   \\ \hline
\end{tabular}

\newpage

{\large{\bf Table 3}}

\vskip 1cm

\begin{tabular}{|c|c|c|c|c|} \hline
$k\setminus g$&   0.01 &   0.3 &  1   & 200   \\  \hline
 2 & -0.07  & -2.0  & -7.4 & -53   \\ 
 3 & -0.07  &  1.4  &  9.0 &  256   \\ 
 4 & -0.07  & -0.4  & -2.4 & -35 \\
 5 & -0.07  &  0.2  &  2.2 &  49 \\ 
 6 & -0.07  & -0.1  & -1.0 & -25 \\
 7 & -0.07  &  0.06 &  0.7 &  14 \\
 8 & -0.07  & -0.03 & -0.5 & -19 \\
 9 & -0.07  &  0.02 &  0.3 &  2.6 \\ 
10 & -0.07  & -0.01 & -0.3 & -15 \\  \hline
\end{tabular}

\newpage

{\large{\bf Table 4}}

\vskip 1cm

\begin{tabular}{|c|c|c|c|} \hline
$k$ & $b_k$  & $\bt_k$ &  $\ep_k(\bt)$ \\ \hline
 2  &  0.729 &   0.176 &   -47  \\ 
 3  &  0.611 &   0.590 &    77  \\ 
 4  &  0.755 &   0.231 &   -30  \\
 5  &  0.669 &   0.409 &    22 \\ 
 6  &  0.756 &   0.257 &   -23  \\
 7  &  0.696 &   0.351 &    5.3  \\
 8  &  0.752 &   0.272 &   -18  \\
 9  &  0.710 &   0.328 &   -1.7  \\ 
 10 &  0.748 &   0.282 &   -16 \\
 11 &  0.718 &   0.317 &   -4.8  \\
 12 &  0.743 &   0.289 &   -13 \\ 
 13 &  0.721 &   0.312 &   -6.3  \\
 14 &  0.739 &   0.294 &   -12 \\
 15 &  0.723 &   0.309 &   -7.2  \\
 16 &  0.736 &   0.298 &   -11 \\
 17 &  0.725 &   0.308 &   -7.5  \\ \hline
\end{tabular}

\end{center}

\newpage

\begin{figure}[h]
\centerline{\psfig{file=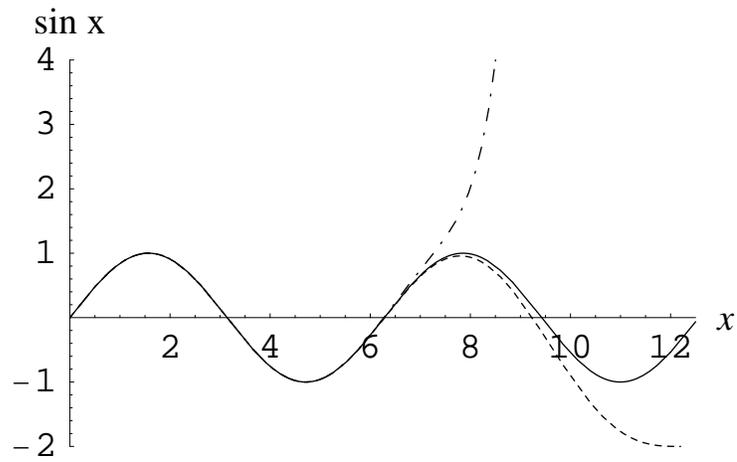,angle=0}}
\vskip 1cm
\caption{Factor approximant $f_{18}^*(x)$ (dashed line) and the
Taylor series of the same order $k=18$ (dashed-dotted line), as compared 
to the exact $\sin x $ (solid line).}
\label{fig:Fig.1}
\end{figure}

\newpage

\begin{figure}[h]
\centerline{\psfig{file=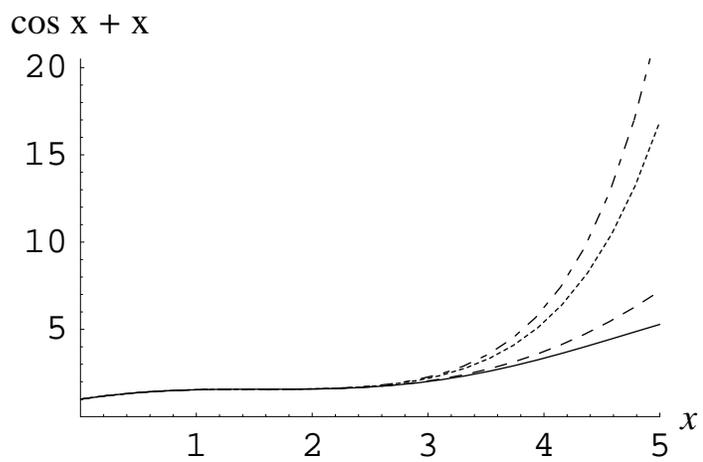,angle=0}}
\vskip 1cm
\caption{
Factor approximants $f_{8}^*(x)$ (dashed-dotted line),
$f_{9}^*(x)$ (dotted line), and $f_{10}^*(x)$ (dashed line), as compared
to the exact function (18) (solid line).
}
\label{fig:Fig.2}
\end{figure}

\newpage

\begin{figure}[h]
\centerline{\psfig{file=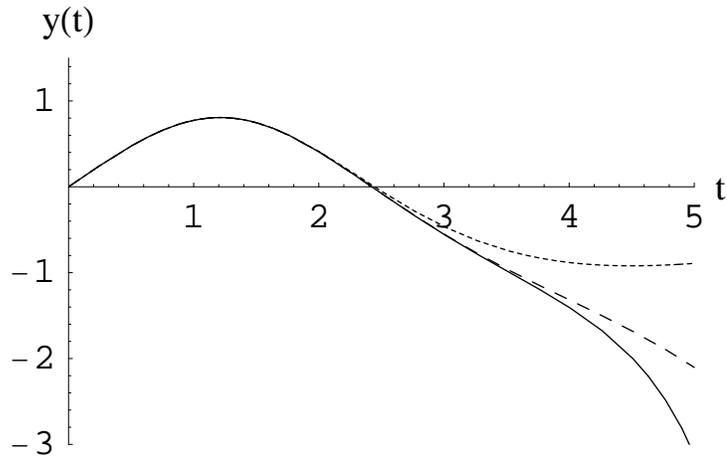,angle=0}}
\vskip 1cm
\caption{
Solutions to the nonlinear oscillator equation (19), with
the initial conditions $y(0)=0$ and $\dot{y}(0)=1$, obtained from the time
series at $t\ra 0$ in the form of the factor approximants $y_k^*(t)$. Here
shown are $y_{14}^*(t)$ (dotted line), $y_{16}^*(t)$ (dashed line), and the
exact numerical solution (solid line).
}
\label{fig:Fig.3}
\end{figure}

\newpage

\begin{figure}[h]
\centerline{\psfig{file=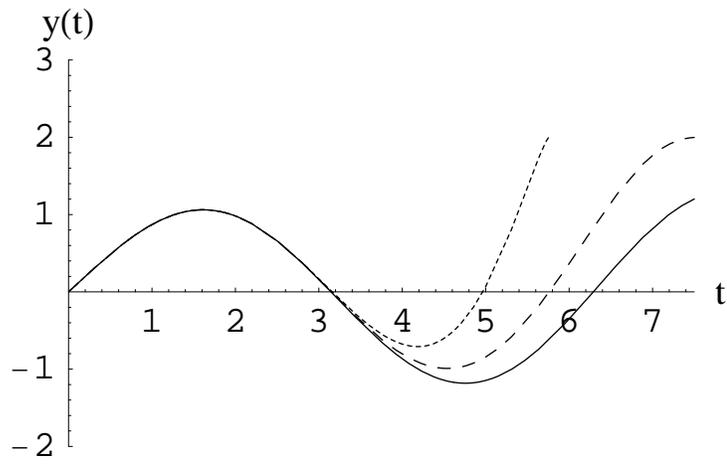,angle=0}}
\vskip 1cm
\caption{
Solutions to the Rayleigh equation (20) in the form of the
factor approximants $y_{18}^*(t)$ (dotted line) and $y_{19}^*(t)$ (dashed
line), compared to the exact numerical solution (solid line).
}
\label{fig:Fig.4}
\end{figure}

\end{document}